\documentclass[aps,pre,twocolumn,showpacs,superscriptaddress,10p,bibliography]{revtex4-2}

\usepackage{amsmath}
\usepackage{amssymb}
\usepackage[sort&compress]{natbib}
\usepackage{graphicx}
\usepackage{color}
\usepackage{physics}
\usepackage{appendix}

\graphicspath{{../Images/}}

\usepackage{float}
\usepackage[caption = false]{subfig}
\begin{document}

\title{Quantum dynamics of a single fluxon in Josephson junctions parallel arrays with large kinetic inductances}
\author{S. S. Seidov}
\affiliation{National University of Science and Technology ``MISIS", Moscow 119049, Russia}
\author{M. V. Fistul}
\affiliation{National University of Science and Technology ``MISIS", Moscow 119049, Russia}
\affiliation{Russian Quantum Center, Skolkovo, Moscow 143025, Russia} 
\affiliation{Theoretische Physik III, Ruhr-Universit\"at Bochum, Bochum 44801, Germany}

\begin{abstract} 
We present a theoretical study of coherent quantum dynamics of a single magnetic fluxon (MF) trapped in Josephson junction parallel arrays (JJPAs) with large kinetic inductances. The MF is the topological excitation carrying one quantum of magnetic flux, $\Phi_0$. The MF is quantitatively described as the $2\pi$-kink in the distribution of Josephson phases, and for JJPAs with high kinetic inductances the characteristic length of such distribution ("the size" of MF) is drastically reduced. Characterizing such MFs by the Josephson phases of three consecutive Josephson junctions we analyse the various coherent macroscopic quantum effects in the MF quantum dynamics. In particular, we obtain the MF energy band originating from the coherent quantum tunnelling of a single MF between adjacent cells of JJPAs. The dependencies of the band width $\Delta$ on the Josephson coupling energy $E_J$, charging energy $E_C$ and the inductive energy of a cell $E_L$, are studied in detail. In long linear JJPAs the coherent quantum dynamics of MF demonstrates decaying quantum oscillations with characteristic frequency $f_{qb}=\Delta/h$.  In short annular JJPAs the coherent quantum dynamics of MF displays complex oscillations controlled by the Aharonov-Casher phase $\chi \propto V_g$, where $V_g$ is an externally applied gate voltage. In the  presence of externally applied dc bias, $I$, a weakly incoherent dynamics of quantum MF is realized in the form of macroscopic Bloch oscillations leading to  a  typical "nose" current-voltage characteristics of JJPAs. As ac current with frequency $f$ is applied the current-voltage characteristics  displays 
a set of equidistant current steps at $I_n=2en f$. 
%current steps with the typical frequency $f_{Bl} \simeq I/(2e)$ %occur, and 
%As the charge $Q$ is trapped in the JJPAs of annular form the %Aharonov-Casher effect realizing in the form of strong oscillations %of the MF band width, occurs. 
\end{abstract}

\maketitle

\section{Introduction}
A great attention has been devoted to a study of solitons, i.e. stable spatially distributed macroscopic structures formed in different nonlinear media \cite{dauxois2006physics}. The solitons have been obtained in various complex solid-state, optical, chemical and biological systems \cite{kartashov2011solitons,scott2018solitons}. An interesting example of topological solitons \cite{manton2004topological} are so-called  \textit{magnetic fluxons} (MFs) found in low-dimensional superconducting systems, e.g. two-dimensional Josephson junctions arrays, long Josephson junctions or Josephson junctions parallel arrays \cite{kivshar1989dynamics,ustinov1998solitons}. Such MFs are vortices of persistent superconducting current, each of them carrying one quantum of magnetic flux, $\Phi_0$. 

An ideal experimental platform to study the \textit{classical} dynamics of MFs is Josephson junctions parallel arrays (JJPAs). A single MF can be trapped in such systems and the dynamics of MF is controlled by externally applied current bias. A large amount of fascinating physical effects in the dynamics of MFs has been theoretically predicted and experimentally observed, e.g. dc current induced resonances \cite{kivshar1989dynamics}, the relativistic dynamics of MF \cite{ustinov1998solitons}, bunching of MFs \cite{vernik1996soliton}, the Cherenkov radiation of plasma modes by moving MF \cite{wallraff2000whispering}, ac current induced dynamic metastable states \cite{fistul2000libration}, just to name a few. From  mathematical point of view the classical dynamics of MF is determined by a large set of coupled nonlinear differential equations \cite{ustinov1998solitons} and a single MF is described as $2\pi$-kink in the spatial distribution of Josephson phases \cite{kivshar1989dynamics,ustinov1998solitons}. 

The next question that naturally arises in this field: is it possible to obtain macroscopic coherent quantum-mechanical phenomena in the dynamics of topological magnetic fluxons? Indeed, the incoherent macroscopic quantum tunneling of magnetic vortices has been theoretically analysed \cite{fazio2001quantum} and observed in two-dimensional Josephson junction arrays \cite{van1996quantum},  macroscopic quantum tunneling and energy level quantization have been observed in the dynamics of a single MF trapped in a long Josephson junction \cite{wallraff2003quantum}. However,  \textit{coherent} quantum effects in the dynamics of MFs have not been observed yet. Observation of coherent quantum dynamics of topological MFs is hampered by two severe obstacles: unavoidably  present dissipation and decoherence, and a rather large size of MF formed in JJPAs with low  inductances, not allowing to map the initial many-body problem to the quantum dynamics of a single degree of freedom.  However, an intensive study of various superconducting lumped elements biased in quantum regime, i.e. superconducting qubits, and networks of interacting superconducting qubits, has already resulted in a  substantial reduction of dissipation and decoherence \cite{krantz2019quantum}. A second problem of reducing the size of MF can be solved by replacement of low geometrical inductances with large kinetic inductances that allows to shrink the $2\pi$ kink distribution. 
%and therefore, drastically reduce the size of MF. 
Large kinetic inductances can be implemented in JJPAs by two methods: embedding in each cell of JJPAs series arrays of large Josephson junctions \cite{matveev2002persistent,manucharyan2009fluxonium}, or using disordered superconducting materials \cite{maleeva2018circuit,hazard2019nanowire,astafiev2012coherent}. Recently,  
some coherent quantum-mechanical effects in the dynamics of MFs trapped in the JJPAs with high kinetic inductances were theoretically studied in Ref. \cite{petrescu2018fluxon}.

In this Article we systematically study the \textit{coherent quantum dynamics} of a single MF trapped in JJPAs with large kinetic inductances. The large kinetic inductance is provided by series arrays of large Josephson junctions, and in this case the dynamics of MF is determined by a single degree of freedom. We analyze in detail the MF energy spectrum originating from the macroscopic quantum tunneling in the effective intrinsic Peierls-Nabarro potential, and the coherent quantum oscillations (quantum beats) of MFs occurring in long linear and short annular JJPAs. We show that the frequency of quantum oscillations obtained in short annular JJPAs can be controlled by the Aharonov-Casher phase \cite{reznik1989question}. In the presence of externally applied dc bias current and taking into account a weak dissipation we obtain Bloch oscillations in the dynamics of a single MF.

The paper is organized as follows: In Section II we present our model for  JJPAs with large kinetic inductances, derive  the total Lagrangian of such a system.  In Sec. III we provide a macroscopic  quantum-mechanical description of the coherent quantum dynamics of a single fluxon trapped in such JJPAs.  For that we elaborate a special approximation where a single fluxon is characterized by Josephson phases of three consecutive Josephson junctions. In Section IV we apply this generic description to analyze in detail various macroscopic quantum phenomena occurring in the dynamics of a single magnetic fluxon, i.e.  the MF energy bands, decaying macroscopic quantum oscillations in long linear JJPAs,  macroscopic quantum beats controlled by the Aharonov-Cashier phase in short annular JJPAs. In Section V a weakly incoherent quantum dynamics of MF in the presence of dc and ac bias currents is discussed. The Section VI provides conclusions. 

\section{JJPAs with large kinetic inductances: Models and Lagrangian}

\begin{figure}[h]
\subfloat[a)]{\includegraphics[width=0.7\linewidth]{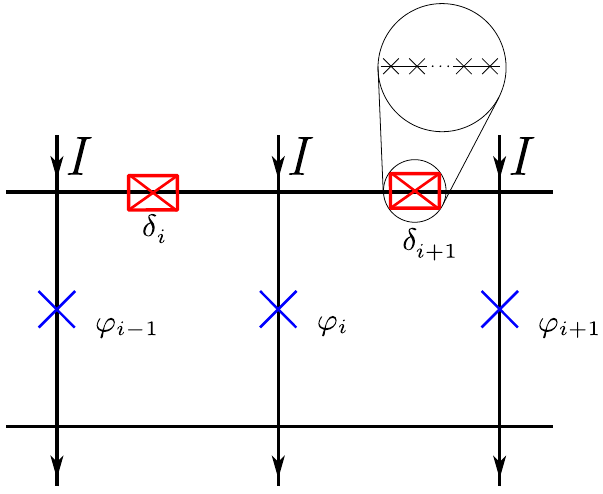}}\\
\subfloat[b)]{\includegraphics[width=0.7\linewidth]{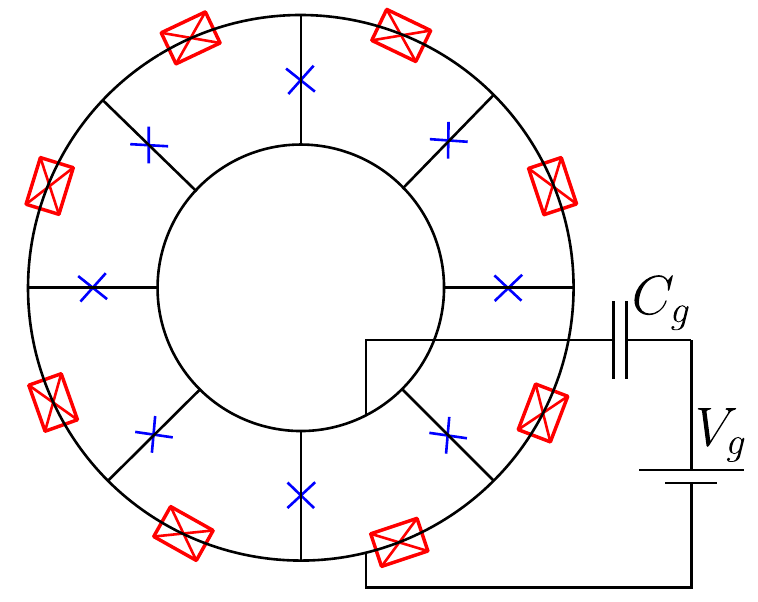}}
\caption{The schematics of JJPAs with large kinetic inductances: a) a linear JJPA; b) an annular JJPA. The bias dc current $I$ and the gate voltage, $V_g$ together with the capaciatance $C_g$ are shown. The Josephson phases of small ($\varphi_i$), and large ($\delta_i$) Josephson junctions are indicated. }
\label{fig1}
\end{figure}

We consider JJPAs with large kinetic inductance composed of $M$ superconducting cells coupled by small Josephson junctions (it is indicated in Fig. \ref{fig1} by blue crosses). The classical dynamics of such JJPAs is determined by a set of time-dependent Josephson phases, $\varphi_i(t)$. In the presence of a single MF trapped in a JJPA, the Josephson phases vary from zero to $2\pi$ on the whole length of JJPA. In order to observe the quantum-mechanical effects in the MF dynamics, the parameters of small Josephson junctions have to be chosen as 
$E_J \geq E_C$, where $E_J$ and $E_C$ are the Josephson coupling energy and the charging energy, accordingly. A high kinetic inductance of JJPAs is provided by  embedding of series arrays of $N$ large Josephson junctions in the upper branch of each cell  \cite{matveev2002persistent,manucharyan2009fluxonium} (it is indicated in Fig. \ref{fig1} by red boxes). 
The Josephson coupling energy $E_{Ja}$, and the charging energy $E_{Ca}$ of these large Josephson junctions were chosen as $E_{Ja} \gg E_{Ca}$ in order to suppress the quantum phase slips in series arrays. The  dynamics of Josephson junctions built in series arrays  is characterized by the time-dependent Josephson phases, $\delta_i$. Each cell is pierced by an externally applied magnetic flux $\Phi_{i}$, and the dc current $I$ is applied in each node. The schematics of various JJPAs are presented in Fig. \ref{fig1}a (the linear JJPA) and \ref{fig1}b (the annular JJPA). Notice here, that in JJPAs of annular form (see, Fig. \ref{fig1}b) , the quantum dynamics of MF can be controlled by an externally applied gate voltage $V_g$ inducing an additional charge on the central superconducting island. Here, $C_g$  is the gate capacitance. 

By making use of the Kirhhoff's circuit laws we write the Lagrangian of the system in the following form \cite{Vool}:
\begin{equation}
\begin{aligned}
& L = K[\dot \varphi_i;\dot \delta_i]-U[\varphi_i; \delta_i]=\\
&=\sum_{i=1}^{M}  \frac{E_J \left (\dot\varphi_i-\frac{2eC_g}{\hbar C}V_g \right)^2}{2\omega_p^2}+
\frac{E_{Ja} \dot\delta_i^2}{2\omega_{pa}^2}  - \\
&-E_J(1 - \cos\varphi_i) -E_J \frac{I}{I_c}\varphi_i -N E_{Ja}(1 - \cos\delta_i), \\
%&g_i = \cos\frac{\Delta_i + \delta_i}{2}.
\end{aligned}
\label{eq:L}
\end{equation}
where $\omega_{p}=\sqrt{8E_JE_C}/\hbar$ and $\omega_{pa}=\sqrt{8E_{Ja}E_{Ca}}/\hbar$ are the plasma frequencies of small and large Josephson junctions, accordingly; $I_c$ is the critical current of a Josephson junction; $C$ is the capacitance of small Josephson junctions. The magnetic flux quantization in each cell leads to a set of constraints on the Josephson phases $\varphi_i$ and $\delta_i$ \cite{Schmidt}:
\begin{equation}
%&\sum_{i=1}^{\mathbf{N}} \delta_i = 0\\ 
%&\sum_{i=1}^{\mathbf{N}} \Delta_i = \frac{2\pi}{N} %\sum_{i=1}^{\mathbf{N}} (n_i + \Phi_i)\\
N \delta_i + \varphi_{i+1}  - \varphi_{i} = 2\pi\left [n_i+\frac{\Phi_i}{\Phi_0} \right ],~~~i=1,....M,  
\end{equation} 
where $\Phi_0$ is the magnetic flux quantum, and $n_i$ is a number of  magnetic flux quanta penetrating the $i$-th cell.  Using such constraints and excluding the phases $\delta_i$ from (\ref{eq:L}) we obtain for the potential energy $U$ the following expression: 
%t allows to express the potential energy through $\varphi_i$:
\begin{equation}
\begin{aligned}
&U\qty(\{\varphi_i\}) = E_J \sum_{i=1}^{M}(1 - \cos\varphi_i)+E_J \frac{I}{I_c}\varphi_i \\
&+ NE_{Ja} \sum_{i=1}^{M} \qty(1 -  \cos\qty[\frac{\varphi_i - \varphi_{i+1}}{N} + \frac{2\pi(n_i + \Phi_i)}{\Phi_0 N}]).
%&g_i = \cos\frac{\Delta_i + \delta_i}{2}.
\end{aligned}
\label{PotentialEnergy}
\end{equation}
%As $E_{Ja} \gg E_{Ca}$ and $E_C \gg E_{Ca}$  the dynamics of large %Josephson junctions is strongly suppressed, 
As $N \gg 1$ expanding the second term in Eq. (\ref{PotentialEnergy}) up to the second order in $1/N$ we obtain the potential energy 
\begin{equation}\label{eq:U}
\begin{aligned}
&U\qty(\{\varphi_i\}) = E_J \sum_{i=1}^{M}(1 - \cos\varphi_i) + E_J \frac{I}{I_c}\varphi_i\\
&+ E_L \sum_{i=1}^{M} \qty(\varphi_i - \varphi_{i+1} + 2\pi \frac{(n_i + \Phi_i)}{\Phi_0})^2, 
\end{aligned}
\end{equation}
where the inductive energy $E_L=E_{Ja}/(2N)$. 

\section{Quantum dynamics of a single MF trapped in a highly inductive ($E_J \gg E_L$) JJPA. }
\begin{figure}[h!!] 
\center\includegraphics[width = \linewidth]{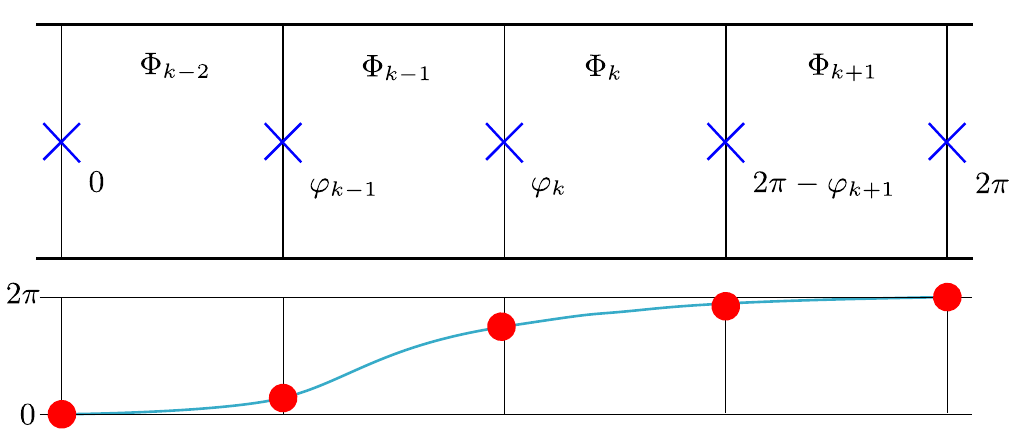}
\caption{The highly inductive JJPA with a single trapped MF (a) and the Josephson phase distribution of a small size MF (b). }
\label{fig2}
\end{figure}

Next, we study a particular case,  $E_{Ja}  \gg E_J$ and $E_C \ll E_{Ca}$, as the quantum dynamics of the Josephson junctions built in series arrays is strongly suppressed. It results in the absence of both the plasma oscillations and macroscopic quantum tunneling (quantum phase slips) in Josephson junctions of series arrays, and therefore, one can neglect the charging energies of series arrays of Josephson junctions in (\ref{eq:L}), and set all $n_i$ to zero in (\ref{eq:U}). 
%throughtrapped magnetic flux quanta in JJPAs, i.e. all $n_i=0$ in %, %and one 
In arbitrary JJPAs a single trapped MF is described as $2\pi$-kink in the distribution of Josephson junctions phases, $\varphi_i$, and for low inductive JJPAs ($E_J \ll E_L$) such distribution spreads over many cells. Here, we assume that JJPAs are highly inductive ones, i.e. $E_J \gg E_L$, and the spatial distribution of Josephson phases becomes a  sharp one. In this case we use a particular approach as a single MF is characterized by the Josephson phases of three consecutive Josephson junctions, and other Josephson phases close to zero or $2\pi$ \cite{BraunFKModel,joos1982properties,furuya1987soliton}. More precisely, we present the MF as a particular \textit{static} Josephson phase configuration: $\{\varphi_i\}_\text{MF} = \{0, \ldots, \varphi_{k-1}, \varphi_k, 2\pi - \varphi_{k+1}, \ldots, 2\pi\}$ where the Josephson phases $\varphi_{k \pm 1}$ are small. Such distribution is schematically presented in Fig. \ref{fig2}.

\begin{figure}[h!!]
\center\includegraphics[width = \linewidth]{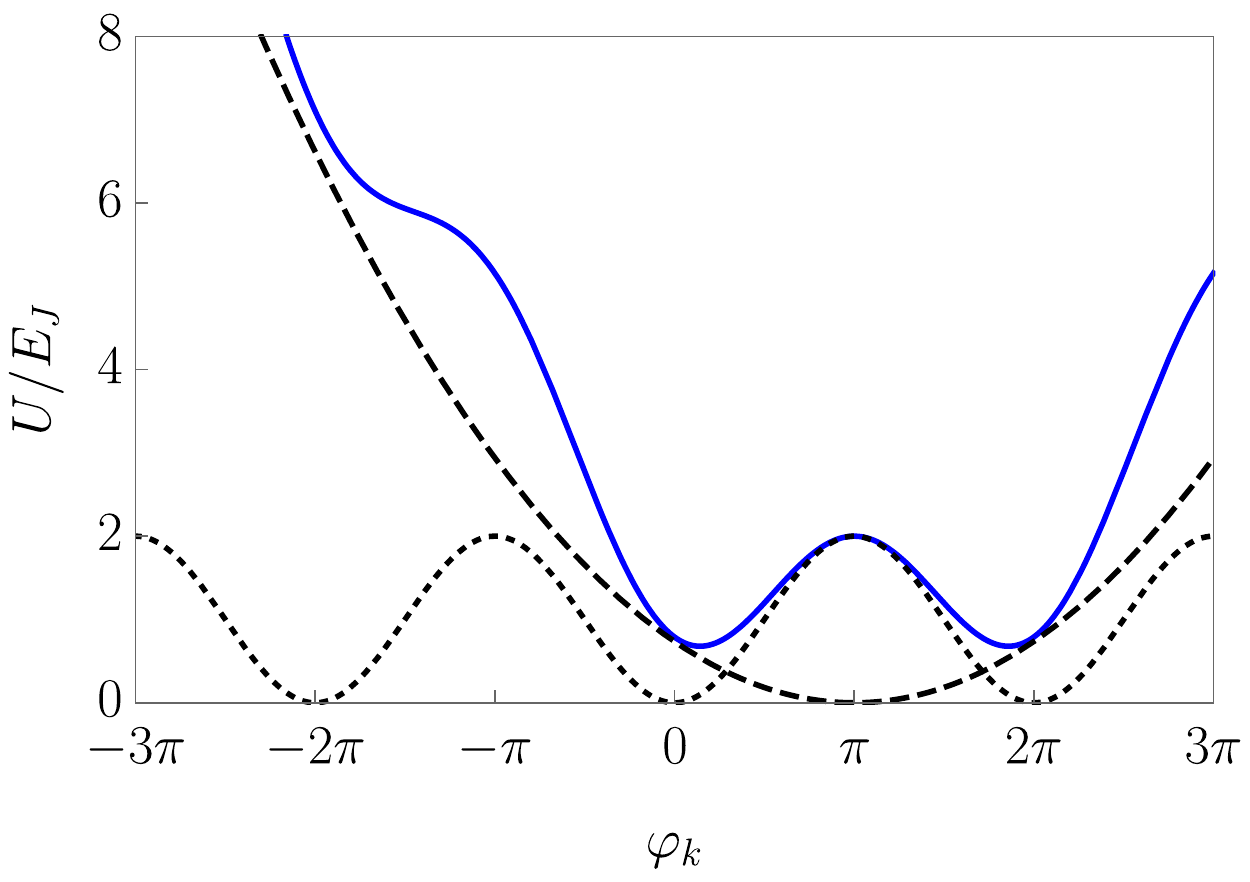}
\caption{The effective potential energy $U_\text{eff}(\varphi_k)$ (solid blue line) for $\Phi_{k-1} - \Phi_k = 0$. Here, the parameter $E_L = 0.04E_J$ was chosen. Dashed and dotted lines represent parabolic- and $\cos$- terms in the Eq. (\ref{EffPotential}), respectively.} \label{3}
\end{figure}

Substituting such Josephson phase configuration in  (\ref{eq:U}) we obtain the effective potential energy $U_\text{eff}$ as follows (here, we consider a specific case $I=0$): 
%%%%%%%%%%%%HERE END!!%%%%%%%%%%%%%%%%%%%%%%%%%%%%%%%%%%%%%%%5
\begin{equation}
\begin{aligned}
&U_\text{eff}[\{\varphi_i\}_\text{MF}] = E_J(3 - \cos\varphi_{k-1} - \cos\varphi_k - \cos[\varphi_{k+1}]) + \\
& +E_L\left[\qty(-\varphi_{k-1} + 2\pi \frac{\Phi_{k-2}}{\Phi_0})^2 +
\qty(-\varphi_{k+1} + 2\pi \frac{\Phi_{k+1}}{\Phi_0})^2+\right. \\ 
&+ \qty(\varphi_{k-1} - \varphi_k + 2\pi \frac{\Phi_{k-1}}{\Phi_0})^2 + \\
&+\left. \qty(\varphi_{k} - 2\pi + \varphi_{k+1} + 2\pi \frac{\Phi_k}{\Phi_0})^2  \right].\\
%&+
\end{aligned}
\end{equation}  
Since we are interested in the low-frequency dynamics of Josephson phases, and the Josephson phases $\varphi_{k \pm 1}$ display the high frequency dynamics only, one can expand the potential $U_\text{eff}$ up to second order in $\varphi_{k \pm 1}$ and minimize the potential energy with respect to them. Using this procedure we write the effective potential $U_\text{eff}(\varphi_k)$ depending on a single macroscopic degree of freedom $\varphi_k$, in the following form:
\begin{equation}
\begin{aligned}
&U_\text{eff}(\varphi_k) =  E_J (1 - \cos\varphi_k)+ \\
&+2E_L\bigg(\varphi_k - \left. \pi \left[1 + \frac{\Phi_{k-1} - \Phi_k}{\Phi_0}\right] \right)^2 
\end{aligned} \label{EffPotential}
\end{equation}
Here for simplicity we set the externally applied magnetic fluxes $\Phi_{k-2}$ and  $\Phi_{k+1}$ to zero, and the condition, $E_J \gg E_L$ is used. The magnetic fluxes $\Phi_{k-1}$ and $\Phi_k$ allow one to control the positions and the relative depths of potential minimums, and for a most relevant case as $\Phi_{k-1} - \Phi_k = 0$, the dependence of $U_\text{eff}(\varphi_k)$ is presented in Fig. \ref{3}. 
In this analysis we neglect the interaction of MF with plasma oscillations in tails of MF, i.e. excitations of $\varphi_{k-1}$ and $\varphi_{k+1}$. This assumption is valid because the strength of such interaction decreases with $E_L$, and it becomes rather small for typical JJPAs with high kinetic inductances.

%
%This gives the potential as a function of $\varphi_k$ and four %parameters $\Phi_{k-2}, \Phi_{k-1}, \Phi_k, \Phi_{k+1}$. %Calculations show, that $\Phi_{k-1}$ and $\Phi_k$ are enough to %control the potentials form, so we set for simplicity. The potential %is 
%Changing one can change the position of the parabolas minimum and %obtain different potential well configurations.

\begin{figure}[h!!]
\center\includegraphics[width = \linewidth]{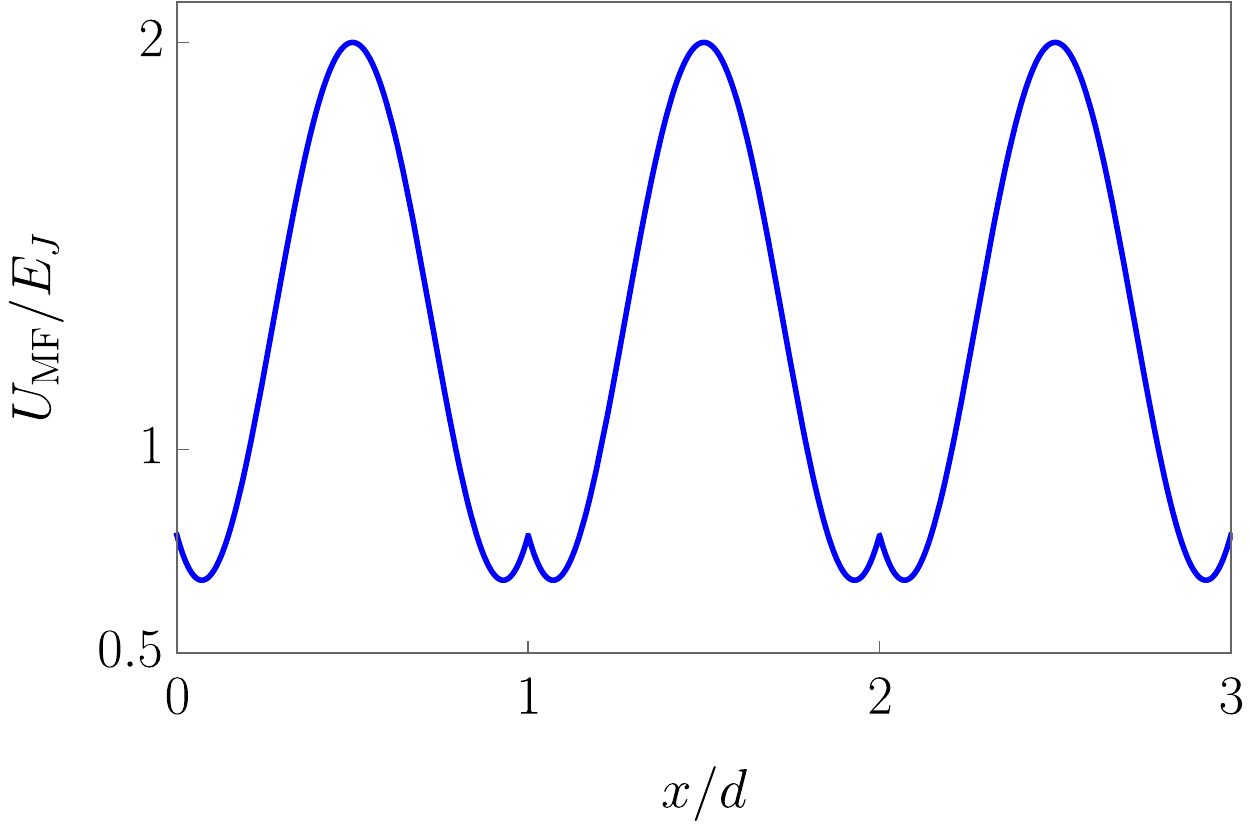}
\caption{Spatial dependence of the potential $U_\text{MF}(x)$ (Eq. (\ref{MFPotential1})). The parameter $E_L = 0.04E_J$ was chosen.
%Black dashed lines denote periodically extended potential $U(x)$.
}
\label{fig:4}
\end{figure}

%\subsection{Quantum dynamics of a single fluxon}
Next, we construct the effective potential energy describing the motion of a single MF along a JJPA. For that, we use the potential (\ref{EffPotential}) on  the Josephson phase segment of $0 \leq \varphi_k \leq 2\pi$. Such potential describes a single MF located in $(k-1)$-th ($\varphi_k \simeq  2\pi$ ) or $k$-th cell ($\varphi_k \simeq 0$ ) of the JJPA. Changing the variable as $\varphi_k = 2\pi-2 \pi x/d$, where $d$ is  the size of a single cell, and the origin of $x$-axis is located in the center of $(k-1)$-th cell, we write down the $U_\text{eff}(x)$ on the interval $0 \leq x \leq d$
\begin{equation}
\begin{aligned}
U_\text{eff}(x) &= E_J\qty(1 - \cos\frac{2\pi x}{d})+2E_L\qty(\frac{2\pi x}{d} - \pi)^2. 
\end{aligned}
\end{equation}
Here, we set the magnetic fluxes as $\Phi_{k-1} - \Phi_k = 0$.
Expanding the $U_\text{eff}(x)$ in Fourier series we obtain the potential energy of a single MF which is valid on a whole axis $x$: 
%Extending periodically this potential to a whole JJPA we obtain the %spatial dependence of a single MF potential energy, $U_{MF}(x)$, as 
%\begin{equation}
%\begin{aligned}
%&\tilde U_{MF}(x) = \sum_{n = 0}^\infty a_n \cos\frac{2\pi n x}{d}\\
%&a_n = \frac{2}{d}\int\limits_0^d U(x)\cos\frac{2\pi n x}{d} \dd x.
%\end{aligned}
%\end{equation}
%Outside of the segment $[0, d]$ the potential $\tilde U(x)$ is a %periodic function:
\begin{equation}
\begin{aligned}
&U_\text{MF}(x) = \frac{2 E_L \pi^2}{3} + E_L \sum_{n=1}^\infty \frac{8}{n^2}\cos\frac{2\pi n x}{d} + \\
&+ E_J\left(1- \cos\frac{2\pi x}{d} \right).
\end{aligned}\label{MFPotential1}
\end{equation} 
The sum can be also expressed as  
%, which gives the potential for any value of $x$, not restricted on segment $[0, d]$:
\begin{equation}
\begin{aligned}
&U_\text{MF}(x) = 2E_L\qty(\frac{\pi^2}{3} + 2 \operatorname{Li}_2[e^{i\frac{2\pi x}{d}}] + 2 \operatorname{Li}_2[e^{-i\frac{2\pi x}{d}}]) +\\
&+ E_J \qty(1 - \cos\frac{2\pi x}{d}),
\end{aligned} \label{MFPotential2}
\end{equation}
where
\begin{equation}
\operatorname{Li}_s (z) = \sum_{k=1}^\infty \frac{z^k}{k^s}
\end{equation}
is the polylogarithm function \cite{Abramowitz}. The potential $U_\text{MF}(x)$ is presented in Fig. \ref{fig:4}.

The same procedure was used for a single MF in the presence of charging energy, i.e. the kinetic energy term in (\ref{eq:L}), and externally applied current $I$ in order to obtain the Lagrangian of a single MF trapped in a JJPA
%introduced to the system via
%\begin{equation}
%j \sum_{k} \varphi_k
%\end{equation}
%term in the Lagrangian. Repeating the calculations above with this %additional term we end up with the Lagrangian
\begin{equation}\label{MFLagr}
\begin{aligned}
&L = \frac{E_J (2\pi)^2}{2\omega^2_p d^2}(\dot x-\alpha V_g)^2 - \frac{2 E_L \pi^2}{3} - E_L\sum_{n=1}^\infty \frac{8}{n^2}\cos\frac{2\pi n x}{d} -\\
&- E_J\qty(1 - \cos\frac{2\pi x}{d}) - E_J \frac{I}{I_c} \frac{2\pi x}{d},
\end{aligned}
\end{equation}
where $\alpha=edC_g/(\pi \hbar C)$. 

\section{Macroscopic quantum effects in the coherent dynamics of a single MF}
Introducing the operator of momentum as $\hat{P}=-i\hbar \dd/\dd x$ we write the Hamiltonian of a single trapped MF as
\begin{equation}
\begin{aligned}
&\hat H = \frac{(\hat{P}+m\alpha V_g)^2}{2m} +  \frac{2 E_L \pi^2}{3} + E_L \sum_{n=1}^\infty \frac{8}{n^2}\cos\frac{2\pi n \hat x}{d} +\\
&+ E_J\qty(1 - \cos\frac{2\pi \hat x}{d}) + E_J \frac{I}{I_c} \frac{2\pi \hat x}{d},
\end{aligned}\label{Hamiltonian}
\end{equation}
where we define the effective mass of MF, $m=E_J (2\pi)^2/(\omega_p d)^2$.

\subsection{Energy bands}
\begin{figure}[h!!]
\center\includegraphics[width = \linewidth]{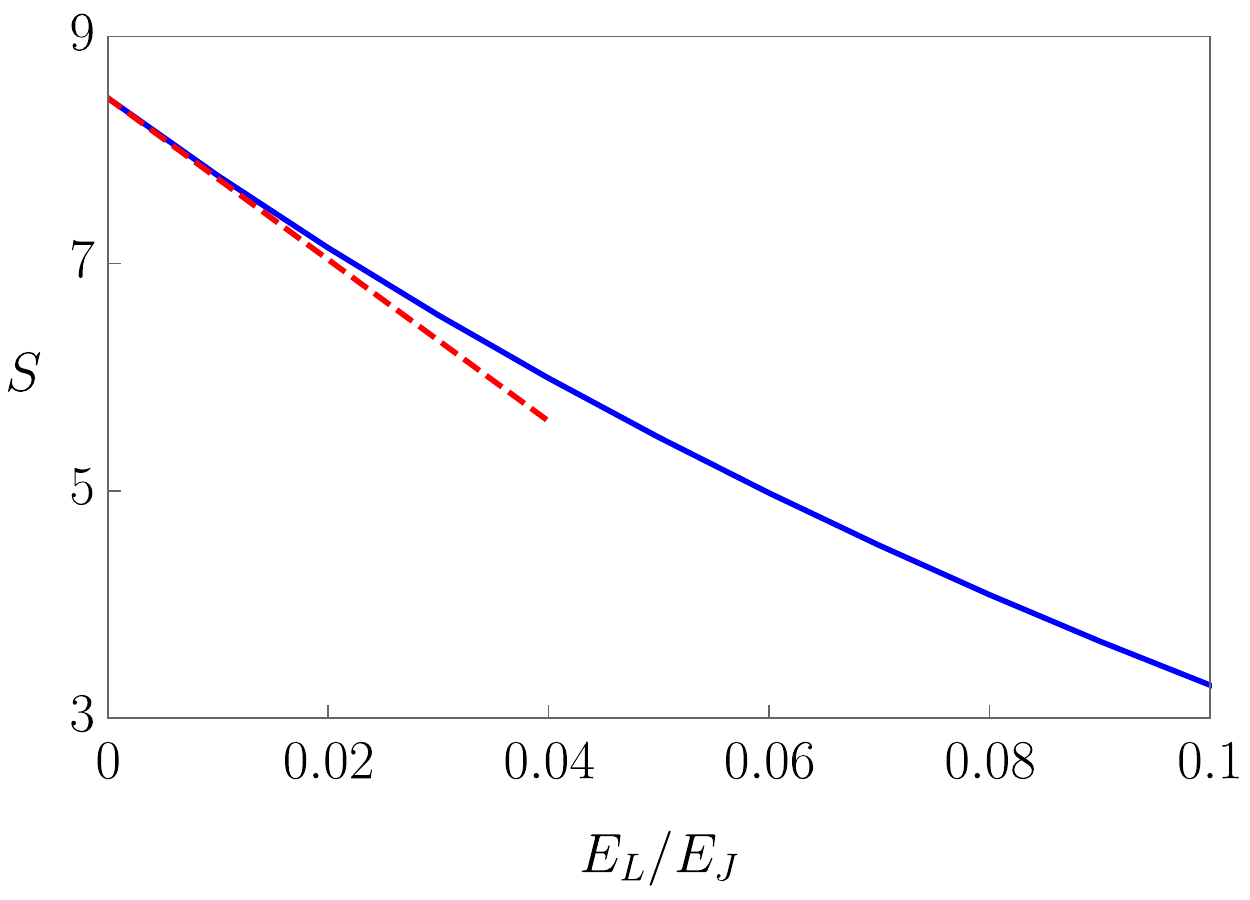}
\caption{The dependence of the lowest band width $S=\ln[2\Delta/(\hbar \omega)]$  on the JJPAs dimensionless inductive energy, $E_L/E_J$.  Solid line --- exact numerical result, dashed line --- linear approximation (\ref{eq:Delta_linear}). The parameters are chosen as $E_C = 0.1E_J$, $E_J/(\hbar \omega_p)=1$.}
\label{Pic5}
\end{figure}

In the absence of both externally applied current $I$ and the gate voltage $V_g$ the coherent quantum dynamics of a single MF is reduced to the quantum dynamics of a single quantum particle moving in the periodic potential $U_\text{MF}(x)$ (\ref{MFPotential2}). It is well known that the eigenfunctions of such quantum problem are determined by the quasi-momentum $p$, and the energy spectrum $E_s(p)$ is composed of infinite number of bands \cite{landau1980statistical}. Moreover, the lowest energy band has a simple form as $E_0(p)=E_0-\Delta \cos (pd/\hbar)$, where $E_0 \simeq E_J$, and the corresponding eigenfunctions are $\psi_p(n)=(1/\sqrt{M})\exp(ipdn/\hbar)$, where $n=0,\pm 1,\pm 2...$ is the cell number of JJPAs. In the limit of $E_J \gg E_C$ the width of energy band $\Delta$ is exponentially small,  and the parameter $\Delta$ determined by tunneling between adjacent  potential wells of $U_\text{MF}(x)$, is obtained in the quasi-classical approximation as \cite{Landau,catelani2011relaxation,schon1990quantum} 
%vels  of a single quantum partiThe energy splitting exponent %between levels in two minimums is given by 
\begin{equation}\label{ParameterDelta}
\begin{aligned}
&\Delta = \frac{\hbar \omega_0}{2} \exp [-S]\\
&S=\frac{1}{\hbar}\int\limits_{x_1}^{x_2} \sqrt{2m|U_\text{MF}(x) - U_\text{MF} (x_1)|}\dd x,
\end{aligned}
\end{equation}
where $x_1$ and $x_2$ are minimums of the potential $U_\text{MF}(x)$ on the interval $0<x<d$, and $\omega_0$ is the frequency of small oscillations, $\omega_0 \simeq \omega_p$.   
%\begin{equation}
%m = \frac{E_J (2\pi)^2}{\omega^2_p d^2}
%\end{equation} 
%is the effective mass of the "particle".
Numerically calculating the integral in (\ref{ParameterDelta}) we obtain the dependence of parameter $S=\ln [2\Delta/(\hbar \omega_0)]$ on $\beta_L=E_J/E_L \geq 1$, and this dependence is shown in Fig. \ref{Pic5} (solid line).  
As the parameter $\beta_L$ is extremely large, i.e. $\beta_L \gg 1$,
we obtain in a linear approximation (see details of calculation in Appendix):
\begin{equation}\label{eq:Delta_linear}
S= S_0 \left (1-\frac{7\zeta(3)}{\beta_L} \right),~~S_0=\frac{8E_J}{\hbar \omega_p}
\end{equation}
This dependence is shown in Fig. \ref{Pic5} by dashed line.
%\section{External current}

\subsection{Coherent quantum oscillations of MF in long linear JJPAs}
\begin{figure}[h!!]
\center\includegraphics[width = \linewidth]{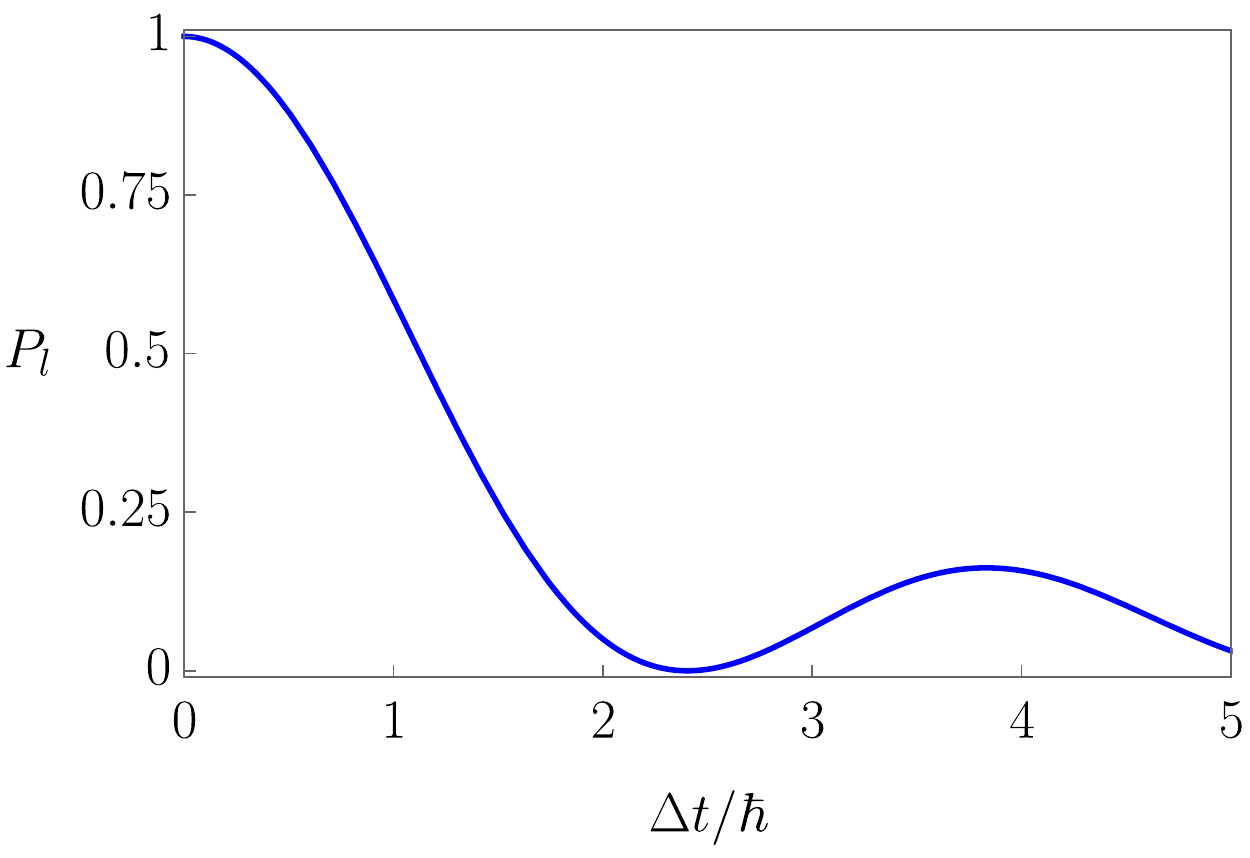}
\caption{The time-dependent probability of $P_l(0,t)$ to obtain quantum MF at the initial position in a long linear JJPA.}\label{fig:wavespread-Long}
\end{figure}

At low temperatures as the excitations to upper energy bands are strongly suppressed, for long linear JJPAs (see, Fig. \ref{fig1}a) the quantum dynamics of MF demonstrates the wave function spreading, and the time-dependent probability
$P_l(x,t)$ to obtain MF at the position with coordinate $x$ is written as
\begin{equation}\label{Probability1}
P_l(x,t) = \left |d \int\limits_{-\infty}^\infty \frac{\dd p}{2\pi \hbar}\exp \left [-\frac{i\Delta}{\hbar} t \cos \frac{pd}{\hbar}-\frac{ipx}{\hbar} \right] \right|^2,
\end{equation}
and e.g. the probability to find the MF in the center of $n$-cell varies with time as $P_l (n,t)=J^2_n (\Delta t/\hbar)$ displaying the decaying quantum beats with the frequency of the order $\hbar/\Delta$. The time-dependent probability to obtain the MF at the initial position, i.e. the center of $0$-th cell, is shown in Fig. \ref{fig:wavespread-Long}. 

\subsection{Coherent Quantum Dynamics of MF in a short annular JJPA: the Aharonov-Casher phase}
Here, we consider the coherent quantum dynamics of MF trapped in a short annular JJPA in the presence of an externally applied gate voltage $V_g$ but the current bias $I$ is still absent. As one can see from the Hamiltonian (\ref{Hamiltonian})  the amplitude of coherent tunneling of MF between the neighbouring cells accumulate the additional phase $\pm \chi$, where $\chi=md\alpha V_g/\hbar$, and the positive (negative) sign corresponds to the tunneling in clockwise (anticlockwise) direction \cite{friedman2002aharonov}. The phase $\chi$ is the seminal Aharonov-Casher phase intensively studied previously in quantum dynamics of magnetic vortices trapped in superconductors or two-dimensional arrays of Josephson junctions \cite{reznik1989question,elion1993observation,van1990aharonov,fazio2001quantum} or various $n$--Josephson junctions SQUIDs \cite{friedman2002aharonov,pop2012experimental}. 

\begin{figure}[h!!]
\subfloat[a)]{\includegraphics[width=\linewidth]{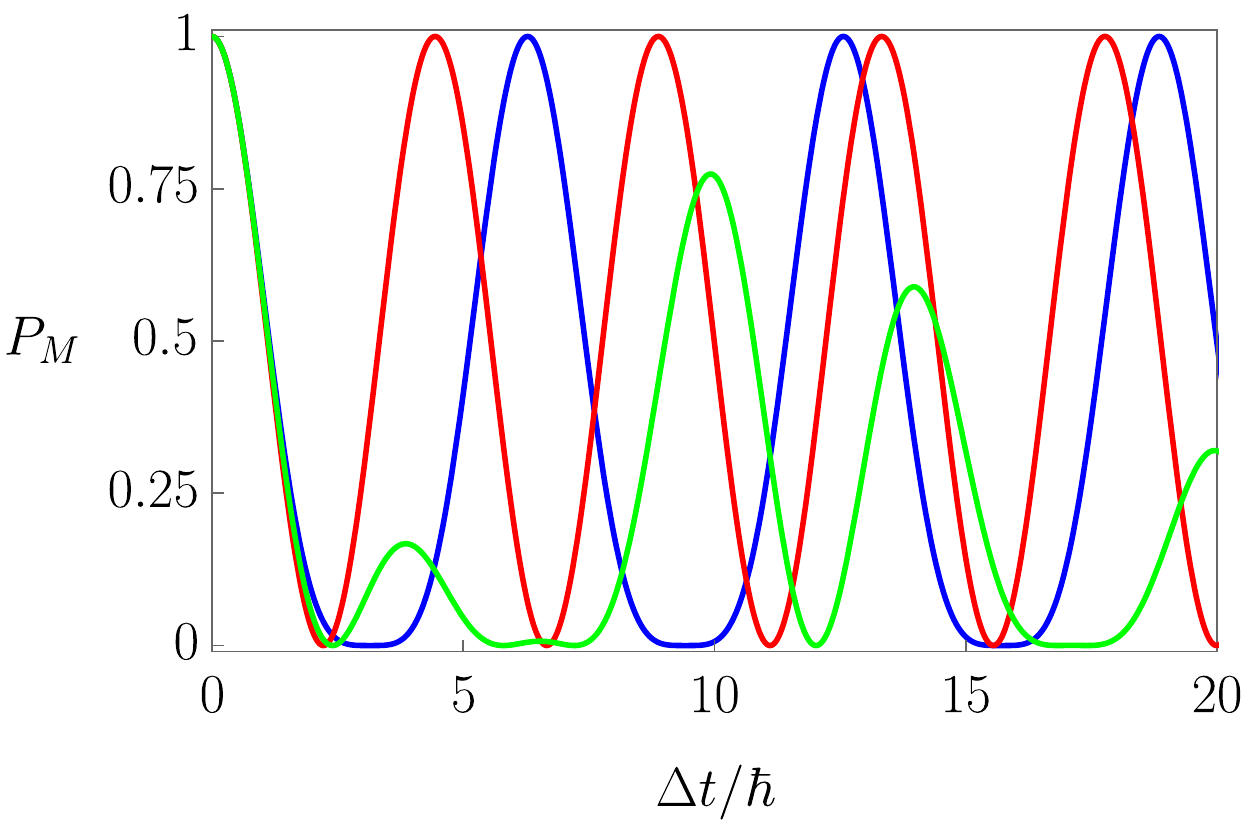}}\\
\subfloat[b)]{\includegraphics[width=\linewidth]{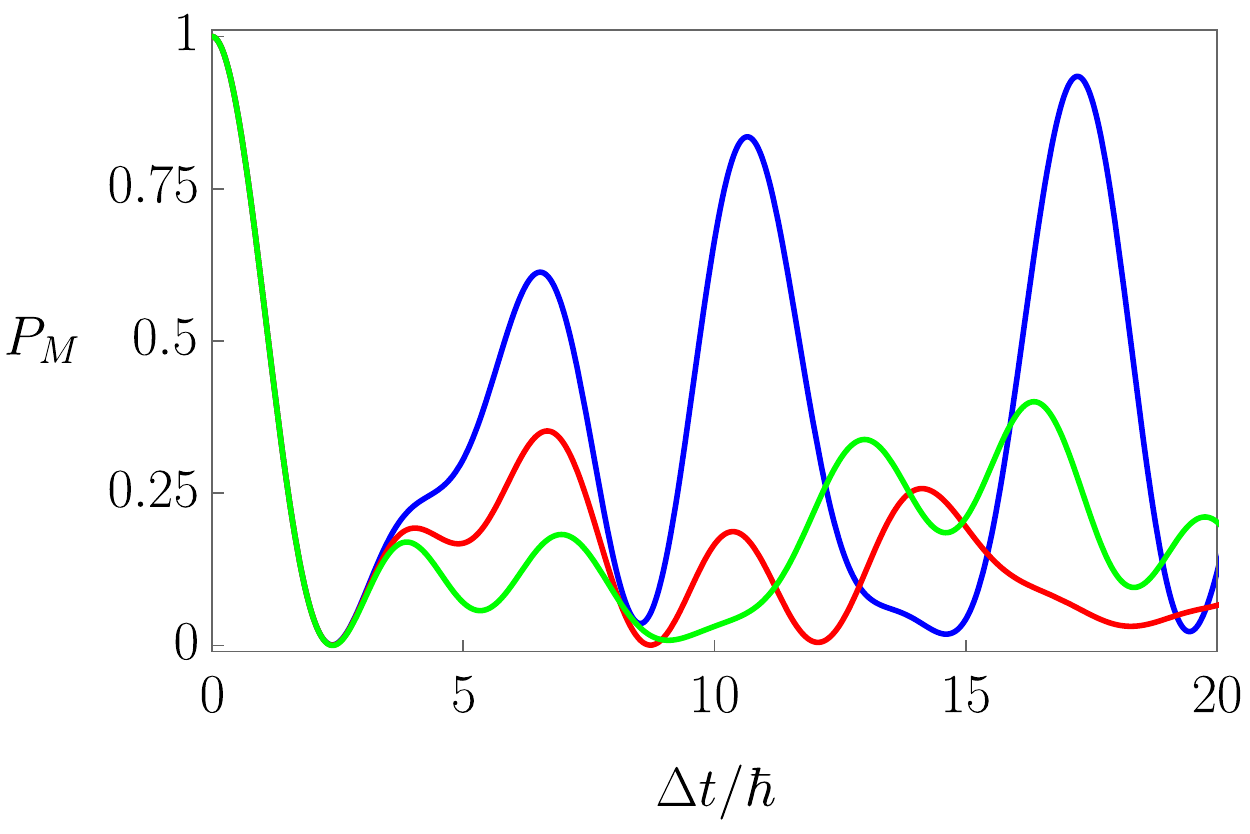}}
\caption{The time-dependent probability of $P_{M}(0,t)$ to obtain the quantum MF at the initial position in  short annular JJPAs of different sizes: $M=4$ (a) and $M=5$ (b) . The values of Aharonov-Casher phase $\chi$ are chosen as $\chi=0 $ (blue lines), $\chi=\pi/8$ (green lines) and $\chi=\pi/4$ (red lines).}\label{fig:wavespread-short}
\end{figure}

For a single MF trapped in an annular JJPA with $M$ cells (see, Fig. \ref{fig1}b)  the eigenfunctions of MF are  $\psi_{p_m}(n)=(1/\sqrt{M}) \exp (ip_m d n/\hbar)$, where $p_m=2\pi m/M$, $m=0,1,2, \ldots, M-1$, and the energy spectrum of the lowest band $E_0(p_m)$ controlled by the Aharonov-Casher phase $\chi$, has a following form:
\begin{equation}\label{Spectrum}
E_0(p_m)=E_0-\Delta \cos (p_md/\hbar+\chi).
\end{equation}
The coherent quantum dynamics of MF in short annular JJPAs of a size $M$ is completely described by probabilities $P_{M}(n,t)$ to obtain  MF in the $n$-cell at time $t$ if MF was initially located in the $0$-th cell. Such probabilities are obtained as follows:
\begin{equation}\label{Probability-short}
P_M(n,t)=\frac{1}{M^2}\left|\sum_{m=0}^{M-1} \exp \left[i\frac{2\pi m n}{M}-i \frac{\Delta t}{\hbar}\cos(\frac{2\pi m}{M}+\chi) \right] \right |^2
\end{equation}
Thus, for $M=2$ we obtain $P_{M=2}(0,t)=\cos^2 [\Delta \cos (\chi)t/\hbar ]$, and therefore, for $\chi=0$ the quantum beats with the frequency of $f_{qb}=\Delta/h$ are realized, but for $\chi=\pi/2$ the quantum beats are completely suppressed, and MF is localized in the $0$--th cell. 
The typical dependencies of $P_{M}(0,t)$ for annular JJPAs of different sizes ($M=4,5$) and a few values of $\chi$ are presented in Fig. \ref{fig:wavespread-short}. Notice here, that for $M=4$ the dependencies 
$P_{M=4}(0,t)$ for $\chi=0$ and $\chi=\pi/2$ accidentally coincide. 

The location of MF and corresponding quantum oscillations for both long and short JJPAs of different geometries can be experimentally verified by the spectroscopy of plasma excitations interacting with a single MF as it was proposed in Ref. \cite{Moskalenko}.

\section{Weakly incoherent quantum dynamics of a single MF}
%\section{Bloch oscillations}
In the presence of externally applied current $I(t)$, the \textit{weakly dissipative dynamics} of a single quantum MF trapped in a  JJPA of large kinetic inductances can be described as follows:
by making use of the quasi-momentum $p$--representation we obtain the operator of MF center $\hat{x}$ as $\hat{x}=-i\hbar \dd /\dd p$. A weak dissipation can be modelled as the interaction of the MF degree of freedom $x$ with the bath of harmonic oscillators $y_i$ \cite{caldeira1983quantum}. The total Hamiltonian is $\hat H_{tot}=\hat H+\hat H_{osc}+\hat H_{int}$, where the interaction Hamiltonian is determined as $\hat{H}_{int}(x, {y_i})=g \hat{x}\sum_i \hat{y}_i$ \cite{caldeira1983quantum,dittrich1998quantum}. In this case the Heisenberg equation of motion for the operator $\dot{\hat{x}}$ is written as 
\begin{equation}\label{Heisenbergequation1}
\dot{\hat x}=\qty[\dv{\hat{p}},\hat{H}]=\dv{E_s(\hat{p}) }{\hat{p}},
\end{equation}
where $E_s(p)$ are the energy bands of the macroscopic quantum particle moving in the potential, $U_\text{MF}(x)$ (see, Eq. (\ref{MFPotential1})). In the presence of dissipation the Heisenberg equation of motion for the operator $\hat{p}$ is written as
\begin{equation}\label{Heisenbergequation2gen}
\dot{ \hat p}=-\frac{1}{\hbar}\qty[\hat{p},\hat{H}]-\frac{1}{\hbar}[\hat{p},\hat{H}_{int}],
\end{equation}
%where the Hamiltonian $\hat{H}_{int}(x, {y_i})=g \hat{x}\sum_i \hat{y}_i$ %represents the interaction of MF degree of freedom $x$ with the bath of %harmonic oscillators, $y_i$. 
For a weakly dissipative case we can trace out the bath degrees of freedom \cite{dittrich1998quantum,likharev1985theory}, and obtain the equation of motion:
\begin{equation}\label{Heisenbergequation2}
\dot{ \hat p}=\frac{2\pi E_J }{dI_c} I(t)-\gamma m\dot{\hat{x}}.
\end{equation}
Here, we introduce the phenomenological parameter $\gamma$ (the inverse relaxation time) characterizing the interaction of the moving topological MF with the environment, i.e. dissipation and decoherence processes. 
Substituting (\ref{Heisenbergequation1}) in (\ref{Heisenbergequation2}), we obtain the dynamic equation for the quantum MF as 
\begin{equation}\label{Heisenbergequation3}
\dot{p}=\frac{2\pi E_J }{dI_c} I(t)-\gamma m \dv{E_s(p)}{p}.
\end{equation}
Since voltage $V$ is determined by the Josephson relationship as $2eV=\hbar \dot{\varphi}=(2\pi \hbar/d)\dot{x}$, and using (\ref{Heisenbergequation1}) we obtain the expression for the voltage as 
\begin{equation}\label{voltage}
V= \frac{\pi \hbar}{ed}\dv{E_s(p)}{p}.
\end{equation}
Notice here that a similar analysis has been carried out long time ago in Refs. \cite{likharev1985theory,schon1990quantum} for small Josephson junctions subject to applied external current $I$. 

\subsubsection{Bloch oscillations and current–voltage characteristics}
First, we analyze the weakly dissipative dynamics of a single MF in the presence of applied  dc current, $I$. The Eq. (\ref{Heisenbergequation3}) has a stationary solution $\dot p = 0$ as $I< I_t$, where
\begin{equation}\label{IT}
I_t=\frac{2\pi \gamma \Delta}{\hbar \omega^2_p} I_c.
\end{equation}
In this regime the voltage $V$ linearly depends on the current $I$ as $V= [\hbar \omega^2_p/(2eI_c\gamma)] I  $. As the current $I>I_t$ the solution $p(t)$ of (\ref{Heisenbergequation3}) is a non-stationary one, and by taking into account  a single lowest energy band we  obtain the solution  analytically  (see details in Appendix B). In this regime the voltage depends periodically on time (see Fig. \ref{fig:PV}), and the period $T$ (see Appendix B) is written as 
\begin{equation}\label{Period}
T =  \frac{2 \pi}{\omega_{0}} \frac{I_t}{\sqrt{I^2-I^2_t}}, ~~I>I_t
\end{equation}
where we introduce the typical frequency of MF Bloch oscillations, $\omega_{0}=(2\pi)^2 E_J \gamma \Delta/(\omega_p \hbar)^2=\pi I_t/e$.  Notice here that the period $T$ increases up to infinity as the current $I$ approaches $I_t$. Substituting the solution of the dynamic equation (see  Eq. (\ref{SolutionDE})) to  (\ref{voltage}) we obtain the voltage $V(t)$ oscillating in time with the frequency, $f_{Bl}=\sqrt{I^2-I^2_t}/(2e)$. The dependence of $V(t)$ displaying  the periodic Bloch oscillations with the frequency $f_{Bl}$, is presented in Fig. \ref{fig:PV} for two different values of $I/I_t$. 
%The voltage is given by
%\begin{equation}
%2eV = \hbar \dot \varphi = \frac{2\pi \hbar}{d} \dot x = \frac{2\pi %\hbar}{d} \dv{E}{p} = \frac{2\pi \hbar}{d} \Delta \sin p(t).
%\end{equation}
%Two functions are plotted in Fig. \ref{fig:PV}.
\begin{figure}[h!!]
\center\includegraphics[width = \linewidth]{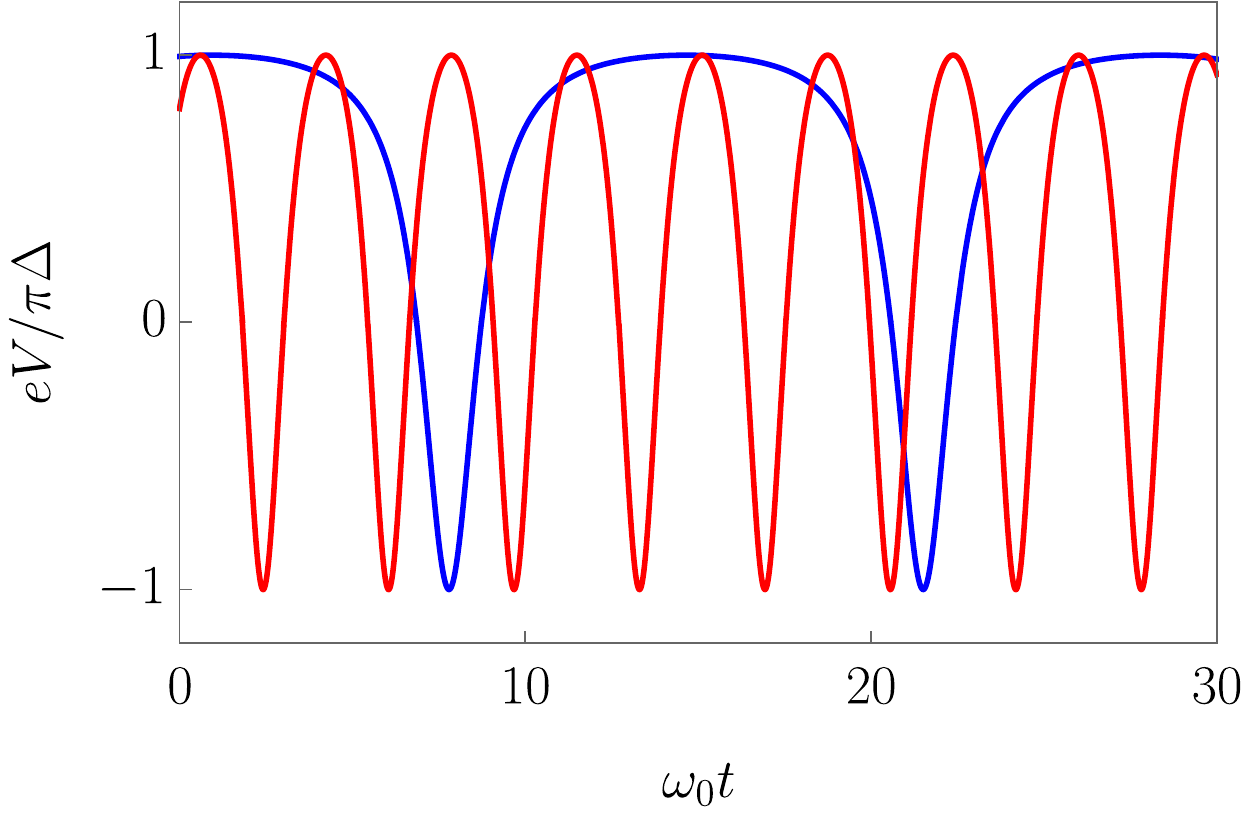}
\caption{The periodic dependence of the voltage, $V(t)$, on time for two values of parameter $I/I_t$, i.e.  $I/I_t=1.1$ (blue line) and $I/I_t=2$ (red line).}
\label{fig:PV}
\end{figure}

Averaging the voltage $V(t)$ over the period allows one to obtain the universal form of the current-voltage characteristics (the $I-V$ curve):
\begin{equation}\label{IVcurve1}
\begin{aligned}
&\ev{V} = \frac{\pi \Delta}{e} \cdot \frac{I-\sqrt{I^2-I^2_t}}{I_t}, ~~I>I_t \\
&\ev{V} = \frac{\pi \Delta}{e} \cdot \frac{I}{I_t}, ~~I<I_t.
\end{aligned}
\end{equation}

\begin{figure}[h!!]
\center\includegraphics[width = \linewidth]{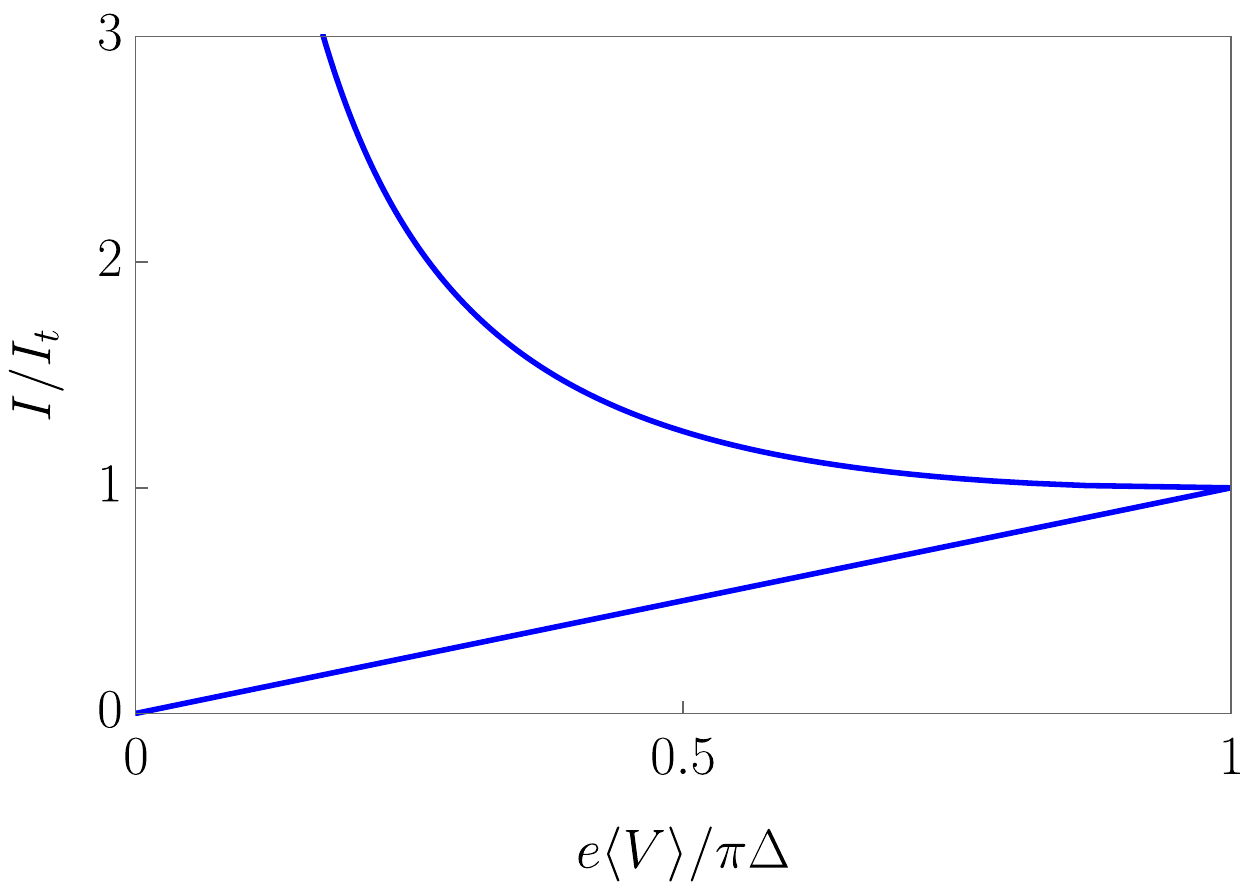}
\caption{The "nose"-type of current-voltage characteristics of JJPAs with a single MF.}
\label{fig:IV}
\end{figure}

%The average voltage for $I/I_c > I_t$ can be calculated using %equation (\ref{Heisenbergequation2}). One has to notice that $V %\sim \dot p$ and $\int_T \dot p \dd t = 2\pi$. Thus we obtain
%\begin{equation}
%\ev{V}_{I/I_c > I_t} = \frac{\pi \hbar \Delta}{e %d}\qty(\frac{I}{I_c I_t} - \frac{d}{E_J I_t T}).
%\end{equation}
Such"nose" type of $I-V$ curve \cite{likharev1985theory,schon1990quantum} is the fingerprint of Bloch oscillations in the dynamics of a quantum MF in a JJPA of large kinetic inductances,  and it  is presented in Fig. \ref{fig:IV}.
In this analysis we neglect the Landau-Zener transitions to upper bands, and since the probability of such transitions is determined as $p_{LZ} \simeq \exp[-\pi (\omega_p)^2 m d/F]$, where $F$ is the slope of the effective potential (the last term in the Eq. (\ref{MFLagr})), this approximation is valid as $I\ll I_c$. In the presence of both dc current $I$ and ac current with the frequency $f$, the seminal current steps located at $I_n=2enf$ can be obtained.
\cite{likharev1985theory,schon1990quantum}.
%\section{Energy level splitting}

\section{Conclusion}
In conclusion, we studied in detail various phenomena occurring in macroscopic quantum dynamics of a topological MF trapped in JJPAs with high kinetic inductances. An implementation of high kinetic inductances in the form of series arrays of large Josephson junctions allows one to drastically reduce the MF size less than the size of a single cell $d$. In this case the quantum dynamics of a single MF is precisely described by a single degree of freedom, i.e. the coordinate of the MF center $x$. By using such description we obtain the MF effective mass $m$, the MF effective kinetic energy controlled by the gate voltage $V_g$, the MF effective potential energy that, in turn, depends on the externally applied current $I$ (see, Eq. (\ref{MFLagr})). In the absence of externally applied current $I$  the effective potential depends periodically on the coordinate $x$, and the energy spectrum of MF contains of infinite number of energy bands. The width of the lowest energy band, $\Delta$, increases with the inductive energy $E_L$ (see, Fig. \ref{Pic5}).  For long linear JJPAs (see the schematic in Fig. \ref{fig1}a) the  parameter $\Delta$ determines the typical frequency of decaying quantum oscillations in the MF wave packet spread (see, Fig. \ref{fig:wavespread-Long}). For short annular JJPAs (see, the schematic in Fig. \ref{fig1}b) the coherent quantum dynamics of MF demonstrates complex quantum oscillations controlled by an external gate voltage $V_g$ through the Aharonov-Casher phase $\chi \propto V_g$ (see, Fig. \ref{fig:wavespread-short}). In particular, we obtain that for a two-cells annular JJPA the frequency of quantum beats is determined as $f_\text{qb}=\Delta \cos(\chi)/h$, and therefore, for $\chi=\pi/2$ the quantum beats are completely suppressed. These quantum oscillations can be experimentally observed through the spectroscopy of plasma oscillations as it was proposed in \cite{Moskalenko}.

As the external dc bias current $I<I_c$ is applied the quantum dynamics of MF displays the seminal Bloch oscillations in the time dependence of the voltage $V(t)$ (see, Fig. \ref{fig:PV}). 
%of the JJPAs with trapped magnetic fluxon . 
These Bloch oscillations results in the "nose"-type $I-V$ curves (Fig. \ref{fig:IV}), and the current steps as both dc and ac currents are applied.

\textbf{Acknowledgements}
We thank Sergej Mukhin and Alexey Ustinov  for valuable discussions.
This work was financially supported by the Russian Science Foundation, Project (19-42-04137). 
 
\appendix

\section{Analytical calculation of the energy band width as $\beta_L \gg 1$}

To obtain the energy band width in the limit of $\beta_L \gg 1$ we expand $S$ up to the first order in $E_L$:
\begin{equation}
\begin{aligned}
& S \approx \frac{1}{\hbar}  \int\limits_0^d \sqrt{2 m E_J \qty(1 - \cos\frac{2\pi x}{d})} \dd x - \\
& - \sqrt\frac{2m}{E_J}\frac{4E_L \pi^2}{\hbar d^2}\int\limits_{x_1}^{x_2} \frac{(x - d + d x_1)(x - d x_1)}{\sqrt{\cos\frac{2\pi x_1}{d} - \cos\frac{2\pi x}{d}}} \dd x.
\end{aligned}\label{eqA1}
\end{equation}
The first integral in (\ref{eqA1}) was calculated exactly:
\begin{equation}
S_0=\int\limits_0^d \sqrt{2 m E_J \qty(1 - \cos\frac{2\pi x}{d})} \dd x = \frac{4\sqrt{m E_J} d}{\pi}=\frac{8E_J}{\hbar \omega_p}.
\end{equation} 
The dependence of $S$ on $E_L$ is written as 
%as a function of $E_L$ is given by
\begin{equation}
S=S_0- \sqrt\frac{2m}{E_J}\frac{4\pi^2}{\hbar d^2} I(E_L),
\end{equation}
where
\begin{equation}
I(E_L) = \int\limits_{\eta E_L}^{d - \eta E_L} \frac{(x - d + d\eta E_L)(x - d \eta E_L)}{\sqrt{\cos\frac{2\pi \eta E_L}{d} - \cos\frac{2\pi x}{d}}} \dd x.
\end{equation}
Here, the limits of integration are written as 
%The limits of integration are functions of $E_L$, so we expand them also up %to first order:
\begin{equation}
\begin{aligned}
&x_1 \approx 0 + \eta E_L\\
&x_2 \approx d - \eta E_L.
\end{aligned}
\end{equation}
The exact expression of $\eta$ is not important here. 
In the first order approximation over $E_L$ one can obtain
\begin{equation}
S=S_0- \sqrt\frac{2m}{E_J}\frac{4\pi^2}{\hbar d^2} I(0),
\end{equation}
and the integral $I(0)$ is calculated explicitly as 
%In the first order the integral can be calculated at $E_L = 0$ giving
\begin{equation}
I(0) = \frac{7 d^3 \zeta(3)}{\sqrt 2 \pi^3} \approx 0.19 d^3,
\end{equation}
where $\zeta(x)$ is the Riemann zeta-function \citep{Abramowitz}.

\section{Current-voltage characteristics of JJPA with a trapped quantum MF}
Introducing the dimensionless variables, i.e. $z=pd/\hbar$ and $\tau=\omega_{0} t$, 
%where the characteristic frequency of Bloch oscillations is  $\omega_{Bl}=(2\pi)^2 E_J \gamma \Delta/(\omega_p \hbar)^2$, 
we obtain the dynamic equation in the following form:
\begin{equation}\label{DimLessEq}
\frac{dz}{d\tau}=\frac{I}{I_t}-\sin(z).
\end{equation}
For $I>I_t$ the solution of Eq. (\ref{DimLessEq}) is written as
\begin{equation}\label{SolutionDE}
z(\tau) = 2\operatorname{Arctan} \left[\frac{\sqrt{a^2-1} \tan \frac{(\tau-\tau_0) \sqrt{a^2-1}}{2}+1}{a} \right ],
\end{equation}
where $a=I/I_t >1 $ and $\tau_0$ is determined by the initial condition. This solution increases with time and periodically oscillates with the dimensionless period $\tilde{T}=2\pi/\sqrt{a^2-1}$. 

%\bibliography{biblio, biblio1}

\end{document}